\documentclass[pre,showpacs,twocolumn,floatfix,amsmath,superscriptaddress]{revtex4}
\usepackage[latin1]{inputenc}
\usepackage[english]{babel}
\usepackage{graphicx,psfrag}
\usepackage{amsmath}
\usepackage{amssymb}
\usepackage{amscd}
\usepackage{eucal}
\usepackage{color}
\usepackage{psfrag}
\usepackage{bm}
\newcommand{\bi}{\bibitem}

\input{epsf}
\begin{document}
\title{Scattering Theory of 
Current-Induced Spin Polarization}
\author{Ph.~Jacquod}
\affiliation{Physics Department,
   University of Arizona, 1118 E. 4$^{\rm th}$ Street, Tucson, AZ 85721, USA}
\date{\today}
\begin{abstract}
We construct a novel scattering theory to investigate 
magnetoelectrically induced 
spin polarizations. Local spin polarizations 
generated by electric currents passing through
a spin-orbit coupled mesoscopic system 
are measured by an external probe. The electrochemical
and spin-dependent chemical potentials on the probe are controllable and
tuned to values ensuring that neither charge nor spin current flow
between the system and the probe, on time-average. 
For the relevant case of a single-channel probe, we find that the resulting
potentials are exactly independent of the transparency of the contact between the probe
and the system. Assuming that 
spin relaxation processes are absent in the probe, we therefore identify the
local spin-dependent 
potentials in the sample at the probe position, and hence the local
current-induced spin polarization, with the spin-dependent potentials
in the probe itself. The statistics of these local chemical
potentials is calculated within random 
matrix theory. While they vanish on spatial and mesoscopic average, 
they exhibit large fluctuations, and we show that single systems typically
have spin polarizations exceeding all known current-induced spin
polarizations by a parametrically large factor. 
Our theory allows 
to calculate quantum correlations between spin polarizations
inside the sample and spin currents flowing out of it.
We show that 
these large polarizations correlate only weakly with spin currents in external leads, 
and that only a fraction of 
them can be converted into a spin current in the linear
regime of transport, which is consistent with the mesoscopic 
universality of spin conductance fluctuations. 
We numerically confirm the theory.
\end{abstract}
\maketitle{}

\section{Introduction}

\subsection{Electric generation of spin polarization}

Two purely electric 
mechanisms have been proposed to generate spin accumulations in
electronic systems with spin-orbit interaction (SOI). 
They are current-induced spin 
polarization~\cite{cisp,Ede90} and the spin Hall effect~\cite{Dya71a,she}. 
Both phenomena appear when a 
spin-orbit coupled conductor is traversed by an electric
current. Current-induced spin polarization manifests itself in a weak
bulk polarization of the electronic spins in a direction determined by the
electric current and the SOI, while the
spin Hall effect manifests itself in
a spin accumulation at the lateral edges of the sample. Though the 
resulting polarizations are usually rather weak,
both effects have been demonstrated 
experimentally~\cite{Kat04,Wun05,Val06,Sai06,Zha06}. They offer
an elegant alternative to spin injection and manipulation with ferromagnets,
Zeeman fields or by optical means, and pave the way toward spintronic
devices with purely electric control of electronic spins. 

A dual aspect of these magnetoelectric effects is that they are 
accompanied by spin currents, 
flowing either inside the sample or leaving it. Early 
analytical investigations of these spin currents have been
devoted to calculations of bulk spin conductivities, but it was soon
realized that the latter are not always related to physically 
relevant quantities.
Local approaches based on classical kinetic equations have been 
used to investigate the conversion of 
magnetoelectric spin accumulations at the edge of a spin-orbit coupled sample
into spin currents leaving the 
sample~\cite{she,Mis04,Bur04,Mal05,Ada07,Tserk}. These quasiclassical theories 
have been quite successful, however
they neglect coherent effects and, in their present form, 
are not appropriate to investigate local
fluctuations and spatial correlations of spin polarization. 
Coherent mesoscopic effects on spin currents 
have recently attracted quite some theoretical attention, both 
analytically~\cite{Cha05,Bar07,Naz07,Kri08,Ada09,Kri09} and 
numerically~\cite{Nik05,guo}, in both diffusive and ballistic systems,
while so far only mesoscopic fluctuations of sample-integrated
spin polarizations in the diffusive regime
have been investigated~\cite{Duc08}. At present, 
little is known about local coherent fluctuations of spin polarizations, 
the scale on which they occur, how they are correlated,
and how they influence spin currents and their 
fluctuations, neither have
current-induced spin polarizations in chaotic ballistic systems 
been investigated so far. 
It is the purpose of this article to start filling these gaps, by
investigating the magnetoelectric generation of spin polarizations
in ballistic mesoscopic samples. 

To that end we construct
a {\it spin-probe model}, 
which connects local spin-dependent chemical potentials to those imposed 
on a nearby probe. In this way, we are able to
express local current-induced spin polarizations in terms of 
transmission coefficients deriving from the scattering matrix. 
Focusing on chaotic ballistic mesoscopic systems, we then use random matrix 
theory ~\cite{Bro96,mehta}
to calculate the statistics of these polarizations. 
We find that, while the ensemble average polarization
vanishes, 
it fluctuates from position to position (in a given sample) and from
sample to sample. For sufficiently large systems, 
the amplitude of these fluctuations by far 
exceeds the magnitude of all
current-induced polarizations investigated to date. 
Simultaneously, we show that this polarization 
is mostly uncorrelated with spin currents in
external leads and thus cannot easily be converted into
a large spin current, at least not in linear response. 


\begin{figure}
\includegraphics[angle=0,width=0.95\columnwidth]{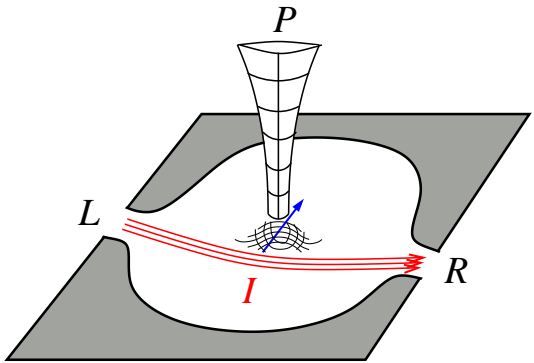}
\caption{\label{fig:probe_sketch} The spin-probe
set-up. An open quantum dot with 
spin-orbit interaction 
is traversed by an electric current ($I$). The magnetoelectrically
generated spin accumulation is measured locally by an external 
probe ($P$) where spin-orbit interaction is absent. 
The charge and spin-resolved electrochemical potentials of the probe
are set such that no spin nor charge current flow between the dot and the
probe. When the probe is contacted to the system via a tunnel barrier of
low transparency, we argue that the spin-dependent 
potentials in the probe can be identified with the local 
spin-resolved potentials
in the sample at the contact with the probe. This allows to relate
spin accumulations
to spin-dependent transmission coefficients.}
\end{figure}

\subsection{The spin-probe model}

Fig.~\ref{fig:probe_sketch} gives a sketch of the system we focus on.
We consider a two-dimensional 
chaotic ballistic quantum dot connected via reflectionless transport
terminals
to two external electron
reservoirs. The reservoirs are biased and an electric current flows through
the dot. 
Assuming that SOI is strong enough that the spin-orbit
time is shorter than the mean electronic dwell time in the dot, 
$\tau_{\rm so} \ll \tau_{\rm D}$, 
spin rotational symmetry is totally broken inside the dot.
Together with chaotic 
boundary scattering, this leads to spin relaxation. 
Does the current flowing through this SOI coupled, spin-relaxing system
generate a spin polarization ?
On one hand, the relative positions of the contacts to the external electrodes
determine an average electronic drift. In that sense, the situation is
very similar to that considered by Edelstein~\cite{Ede90}, and one would 
accordingly argue
that a current-induced spin polarization should emerge.
On the other hand, however, electronic
distribution functions are not well defined locally in such a system
where momentum and energy relax on the scale of the system itself. 
One might
thus wonder how local spin-dependent potentials can be consistently 
defined. 

Here, we circumvent this difficulty by
introducing an additional mobile terminal whose purpose it is to 
probe the local chemical potentials on the dot. 
We assume that spin currents are well
defined in the probe as well as in the transport terminals, which presupposes
that spin relaxation processes are frozen there. In particular, SOI is present
only in the dot. 
The probe is assumed to carry a single-transport channel.
It is connected at one end to a macroscopic external reservoir, which
defines its electrochemical and spin-dependent
chemical potentials unambiguously, while
its other end is contacted to the quantum dot via a tunnel barrier of transparency $0 \le \Gamma_P \le 1$. 
When $\Gamma_P \ll 1$, the probe 
does not perturb the dot and we can identify the 
spin-dependent chemical potentials on the dot
with the corresponding 
probe potentials defined by the condition that neither charge nor spin currents 
flow between the probe and the dot~\cite{caveat1}. 
The probe being mobile, it can
scan the whole dot area, and being single-channeled, it has maximal
resolution. It can thus map the spin-polarization in the
dot with maximal accuracy. 

\subsection{Summary of main results}

The {\it spin-probe model} 
becomes a powerful theoretical tool to calculate spin accumulations once 
electric and spin currents in the probe and the two transport terminals
are related to 
chemical potentials via the scattering approach to 
transport~\cite{But86,Imr86}.
When the probe carries a single transport channel, time-reversal symmetry
and the unitarity of the scattering matrix
ensure that the reflection coefficients from the probe back to itself
are diagonal in spin
indices, independently of $\Gamma_P$. This allows to directly express the electrochemical 
and spin-dependent chemical potentials of the probe
in terms of spin-dependent transmission coefficients from the transport
leads to the probe, and the chemical potentials at those leads. This is done
below in Eq.~(\ref{eq:accu}). We additionally find that,  for a single-channel
probe, the dependence that transmission coefficients 
have on $\Gamma_P$ factorizes. This has the important, 
and somewhat surprising consequence that 
the chemical potentials 
ensuring the vanishing of the currents through the probe, being given
by the ratio of transmission coefficients, do not depend
on $\Gamma_P$. This is an exact result, valid as long as one has a single-channel probe, and we stress
that it is true not only on average nor to some leading order.
Consequently, we are able to 
obtain the value of the spin-dependent potentials on the probe in the
limit $\Gamma_P \rightarrow 0$, where it is justified to
identify the potentials on the dot with those on the probe, while 
performing the
calculation for the analytically simpler case $\Gamma_P=1$. 

Our scattering theory of spin accumulations can be applied to both
diffusive and ballistic mesoscopic systems in the linear response
regime. Because current-induced spin polarization has not been
considered in these systems so far, we specialize here to 
ballistic systems, and accordingly apply random matrix theory
(RMT)~\cite{Bro96,mehta} to calculate average and fluctuations of the 
spin-resolved local chemical potentials. We perform our calculation
for $\Gamma_P=1$ to avoid unnecessary 
complications arising from the presence of
a Poisson kernel for $\Gamma_P < 1$~\cite{Bro96}.
We find that, within RMT, there is no
current-induced local spin polarization on mesoscopic average. However,
sample-dependent or probe position-dependent 
fluctuations make local spin-dependent chemical potentials
fluctuate with a variance
\begin{equation}\label{eq:varmup_N}
{\rm var} \, \mu_P^{(\beta)} \simeq  \frac{(e V)^2}{32 N} \, ,
\end{equation}
given by the voltage bias $V$ between the two transport leads and the
number $N$ of channels each of them carries. The above formula 
gives the leading order contribution in $1/N \ll 1$. 
The exact formula is given below in
Eq.~(\ref{eq:var_mup}). Within RMT, the variance is the same for
$\beta=x,y$ and $z$.

The transmission coefficients from the transport terminals to the
probe are elements of a scattering matrix which depends on
the position
of the probe. Therefore, as the probe scans the
sample, these coefficients fluctuate.
Below, we conjecture and numerically confirm that,
within RMT, moving a single-channel terminal 
by a single Fermi wavelength is sufficient to decorrelate
transmission coefficients to and from that terminal. 
This has for consequence that the total polarization integrated over
the entire dot, 
while still averaging to zero, acquires a sample-to-sample 
variance given by the
single-channel local variance multiplied by the total number 
$k_{\rm F}^2 A/4 \pi$ of uncorrelated probe positions. Thus,
\begin{equation}\label{eq:vartot_N}
{\rm var} \, \mu_{\rm tot}^{(\beta)} 
\simeq \frac{k_{\rm F}^2 A}{4 \pi} \, \frac{(e V)^2}{32 N} \, .
\end{equation} 
Because $\mu_{\rm tot}^{(\beta)}$ gives the 
chemical potential difference between the two possible 
spin orientation along the spin axis $\beta=x,y,z$, dividing it by twice the
level spacing converts it into a number of polarized spins in that direction. 
This gives the typical number of polarized spins in a given sample as
\begin{equation}\label{eq:typaccu}
\langle S^{(\beta)}_{\rm typ} \rangle = 
\frac{{\rm rms} \, \mu_{\rm tot}^{(\beta)}}{2 \delta} 
\simeq \frac{C}{16} \,
\sqrt{\frac{k_{\rm F} L^2}{2 W}}  \, \frac{e V}{\delta} \, , 
\end{equation} 
with the dot level spacing $\delta=2 \pi/ m A$, the width $W$ of the
transport terminals and the linear dot size $L$.
Eq.~(\ref{eq:typaccu}) is one of our
main results. It predicts that linear electric transport across a 
typical, ballistic mesoscopic quantum dot generates
a giant polarization of the electronic spins inside the dot in the
semiclassical limit 
$k_{\rm F} L \gg 1$, and for a large enough dwell time $L/W \gg 1$.
This typical polarization is in particular much larger than the 
average magnetoelectric
polarization predicted by Edelstein~\cite{Ede90}
\begin{equation}\label{eq:Edel}
\langle S^{(\rm E)} \rangle = 4 \frac{k_\alpha \ell}{k_{\rm F} L}
 \, \frac{e V}{\delta} \, ,
\end{equation}
in a diffusive dot with elastic mean free path $\ell \ll L$ and with 
Rashba SOI whose strength $k_\alpha$ is usually only a fraction of $k_{\rm F}$.
It is also much larger than the typical Duckheim-Loss 
polarization~\cite{Duc08},
\begin{eqnarray}\label{eq:DL}
\langle S^{(\rm DL)}_{\rm typ} \rangle  \propto 
\frac{1}{k_{\rm F}\ell \, k_\alpha L} \, \frac{e V}{\delta} \, , 
\end{eqnarray}
predicted for $k_\alpha L \gg 1$.
Despite its magnitude, 
$\langle S^{(\beta)}_{\rm typ} \rangle$ cannot easily 
be converted into a large spin current. Below we 
demonstrate this by calculating the 
correlator of the spin polarization with the spin current in external leads, 
and show that this correlator vanishes unless the probe is within
a Fermi wavelength of the contact to the terminal. In other words, RMT predicts that
only accumulations right at the boundary of the sample can be extracted and
converted into spin currents in the
linear regime, which is consistant with the universal fluctuations of
spin currents found in Refs.~\cite{Bar07,Naz07,Kri08,Ada09}. It is at present
unclear if more spin current can be extracted in nonlinear regimes of transport. 

Having summarized our method and main findings, we now proceed 
to present our theory more formally. In the next Section~\ref{section:scatt}, 
we construct a
scattering theory of current-induced spin polarization. In 
Section~\ref{section:rmt}, we focus on ballistic systems and apply 
RMT to calculate the mean and variance of the local as well as of the
sample-integrated polarizations. We further demonstrate the power of our 
approach by calculating correlators of spin polarizations with spin currents.
Section~\ref{section:numerics} is devoted to a numerical proof-of-principle
of our theory and a quantitative confirmation of our main result,
Eq.~(\ref{eq:typaccu}). We conclude with 
a brief discussion of our results and
of future challenges and demonstrate the $\Gamma_P$-independence of the 
probe voltages for $N_P=1$ in an appendix.

\section{Scattering theory of current-induced spin polarization}\label{section:scatt}

In its standard form, 
the scattering approach to transport gives a linear relation between
currents and voltages measured in external leads~\cite{But86,Imr86}. 
It applies equally well to diffusive and ballistic systems. Extended to take
account of spin currents and chemical potentials in a multiterminal
geometry, the relation reads
\begin{equation}\label{scatt0}
I_i^{(\alpha)} = \frac{e}{h} \sum_{j,\beta}(2N_i \Gamma_i \delta_{\alpha\beta} \delta_{ij}-\mathcal{T}^{(\alpha\beta)}_{ij})\mu_{j}^{(\beta)}\, ,
\end{equation}
where Greek letters label spin indices and Latin letters
label electric terminals. 
Here, $N_i$ gives the number of transport channels in
terminal $i$, $\Gamma_i$ the transparency of the contact between that terminal and the dot, 
$\mu_i^{(0)} = E_{\rm F}+e V_i$ gives the electrochemical potential
in terms of the Fermi energy $E_{\rm F}$ and the applied 
voltage $V_i$, and $\mu_i^{(\beta)}$
give the chemical potential difference between the two possible 
spin orientation along the spin axis $\beta=x,y,z$~\cite{caveat2}.
We further defined
$I^{(0)}_i \equiv I_i$ and $I_i^{(\alpha)}$ as the charge current
and the three components of the spin current vector respectively,
and introduced the spin-dependent transmission coefficients~\cite{Ada09}
\begin{equation}\label{tr_coeff_spin}
\mathcal{T}_{ij}^{(\alpha\beta)} =\sum_{m \in i,n \in j}
{\rm Tr} [ t_{mn}^\dagger
 \sigma^{\alpha} t_{mn} \sigma^{\beta}],
\end{equation}
where $\sigma^{\alpha}$, $\alpha = x,y,z$ are Pauli matrices and
$\sigma^{0}$ is the identity matrix.
The trace in Eq.~(\ref{tr_coeff_spin}) 
is taken over the spin degree of freedom and $t_{mn}$ is a 2$\times$2
matrix of spin-dependent transmission amplitudes from channel $n$ in
lead $j$ to channel $m$ in lead $i$. 

We are interested in the mesoscopic counterpart of the 
magnetoelectric geometry of Ref.~\cite{Ede90}, and
consider a quantum dot connected to two external transport terminals
$L$ and $R$.
An electric bias voltage $V$ 
is applied between the terminals,
$\mu_{L}^{(0)}-E_{\rm F}= -\mu_{R}^{(0)}+E_{\rm F} = e V/2$, where we further
assume that there is no spin accumulation,
$\mu_{L,R}^{(\beta)} = 0$, $\beta\ne 0$. To measure local chemical
potentials, we introduce an external probe with controllable
chemical potentials, which scans the dot surface. The situation
is sketched in Fig.~\ref{fig:probe_sketch}.
The chemical potentials in the probe are
determined by the condition that no charge nor spin current flows
through it, i.e. they are solutions of 
\begin{equation}\label{eq:stm}
I_{P}^{(\alpha)} = 
\frac{e}{h} \sum_{j=L,R,P} 
\sum_{\beta}(2 \Gamma_P N_P \delta_{\alpha\beta} \delta_{{\rm P},j} 
-\mathcal{T}^{(\alpha\beta)}_{{\rm P},j})\mu_{j}^{(\beta)} \equiv 0 \, ,
\end{equation}
and depend a priori on the transparency of the contact between the probe
and the dot. Eq.~(\ref{eq:stm})
determines the local chemical potentials $\mu_P^{(\beta)}$
as a function of the electrochemical potentials $\mu_{L,R}^{(0)}$ 
imposed on the longitudinal 
leads. 
We consider a single-channel probe $N_{\rm P}=1$, which not only 
delivers the sharpest resolution of the dot's chemical potentials landscape,
but moreover ensures that the spin
reflection matrix $\mathcal{T}_{PP}^{(\alpha \beta)}$ from the probe 
to itself is
{\it exactly} (i.e. not only on average) 
diagonal in spin indices.
This follows from time-reversal symmetry, current conservation
and gauge invariance~\cite{symmetry_gs0}, which additionally require
$\mathcal{T}_{PP}^{(\alpha \alpha)} = \mathcal{T}_{PP}^{(00)}$, 
$\forall \alpha$.
One then straightforwardly obtains the chemical potentials in the probe as
\begin{equation}\label{eq:accu}
\mu_{P}^{(\beta)} = [\mathcal{T}^{(\beta 0)}_{PL} - 
\mathcal{T}^{(\beta 0)}_{PR} ] e V \Big/ 2 [
\mathcal{T}^{(00)}_{PL} + \mathcal{T}^{(00)}_{PR}].
\end{equation}
In particular, for $N_{L,R} \gg 1$, the electrochemical
potential in the probe is given by
\begin{equation}\label{eq:accu_charge}
\mu_{\rm P}^{(0)} = [\mathcal{T}^{(0 0)}_{PL} - 
\mathcal{T}^{(0 0)}_{PR} ] e V \Big/4 \Gamma_P  \, ,
\end{equation}
which reproduces a result first derived in Ref.~\cite{But89}.
In diffusive samples, this
gives a linear average 
decay of $\mu_P^{(0)}$ from $eV/2$ to $-eV/2$ as the probe is
moved 
across the sample from regions more easily accessible from the left lead
(where $\mathcal{T}^{(0 0)}_{PL} >
\mathcal{T}^{(0 0)}_{PR}$) to regions which are more easily accessible
from the right. 

We next make the key observation that the $\Gamma_P$-dependence of the transmission
coefficients $\mathcal{T}^{(\alpha \beta)}_{PI}$ between the probe and the
transport terminals $I=L,R$ exactly factorizes when the probe is 
single-channeled. This remarkable property is apparently mentioned here for the first time
and is discussed in some detail in Appendix A. Consequently,
the chemical potentials in Eq.~(\ref{eq:accu})
remain the same, regardless of the transparency of the tunnel barrier between
the dot and the probe, when the latter carries only a single channel. 
Our aim being to obtain the current-induced spin polarization and thus the
spin-dependent chemical potentials on the dot, we are quite naturally interested
in the limit $\Gamma \rightarrow 0$, where
the probe does not perturb the dot and accordingly
the spin-dependent chemical potentials on the dot equal those on
the probe. The observation we just made allows us to access that limit
via an analytically simpler calculation at $\Gamma_P=1$ which we present in 
the next section.

To get the total spin polarization, we scan the probe over the whole
sample. The transmission amplitudes $t_{mn}$ in Eq.~(\ref{tr_coeff_spin}) 
are elements of a scattering matrix that depends parametrically on the
probe position. For a wide probe, $N_P \gg 1$, we argue 
semiclassically that the correlator $
 \langle \mathcal{T}^{(\alpha 
\beta)}_{PL} \mathcal{T}^{(\alpha \beta)}_{P'L} \rangle-  \langle \mathcal{T}^{(\alpha 
\beta)}_{PL} \rangle \langle \mathcal{T}^{(\alpha \beta)}_{P'L} \rangle $ 
decays to zero when the distance between $P$ and $P'$ is of order
the width of the probe. This is so, because beyond that, 
correlated escapes of pairs of trajectories no longer contribute~\cite{Jac06}.
We next make the leap of faith that this semiclassical argument
can be extended to the case of present interest of 
a single-channel probe, for which we accordingly conjecture 
that two measurements
performed a distance $\sim k_{\rm F}^{-1}$ apart deliver uncorrelated
outcomes. To the best of our knowledge, this problem has not been investigated
yet, and we will numerically confirm the validity of this conjecture
below. Further assuming a sharp decay of the correlator at $k_{\rm F}^{-1}$, 
we obtain that the sample-integrated chemical potentials are
given by the sum 
$\mu_{\rm tot}^{(\beta)} = \sum_P \mu_P^{(\beta)}$
over $k_{\rm F}^2 A/4 \pi$ uncorrelated probe positions on the surface
of our two-dimensional quantum dot.

\section{Random Matrix Theory}\label{section:rmt}

Having expressed local and sample-integrated polarizations in terms of
transmission coefficients, Eq.~(\ref{eq:accu}), 
we next proceed to evaluate them quantitatively.
We focus on ballistic chaotic cavities, and accordingly 
use RMT to calculate the average and variance of 
$\mu_P^{(\beta)}$ and $\mu_{\rm tot}^{(\beta)}$. This calculation is easier
for $\Gamma_P=1$, since then the ensemble-average scattering matrix 
vanishes~\cite{Bro96}. Because of the factorization property of the
$\Gamma_P$-dependence of transmission coefficients mentioned above, 
the spin-dependent potentials on the dot are equal to those on the probe for
$\Gamma_P \rightarrow 0$, which in their turn remain unchanged 
as $\Gamma_P \rightarrow 1$. Thus we can extract the dot's 
spin-dependent potentials 
$\mu_{x_P}^{(\beta)}$ at point $x_P$ underneath the probe 
from a calculation at $\Gamma_P=1$, because $\mu_{ x_P}^{(\beta)}
= \mu_P^{(\beta)}(\Gamma_P \rightarrow 0) = \mu_P^{(\beta)}(\Gamma_P=1)$. 
Because we are interested in systems with time reversal symmetry but
totally broken spin rotational
symmetry, we take the scattering matrix as an element of the 
circular symplectic ensemble of random matrices~\cite{mehta}. 
RMT does not depend on the exact form of SOI, but only requires that 
SOI totally breaks spin rotational symmetry. Our results are thus 
expected to apply
for Rashba and Dresselhaus SOI equally well, as long as one has a ballistic
chaotic quantum dot with totally broken spin rotational symmetry.

For $N_P=1$, we get from
Ref.~\cite{Ada09}
\begin{equation}
\langle \mathcal{T}_{ij}^{(\alpha \beta)} \rangle_{\rm RMT} 
= \frac{2 \delta_{\alpha \beta} \, [ N_i N_j \delta_{\alpha 0} - 
(\delta_{\alpha 0}-1/2) N_i \delta_{ij}]}{N_{L}
+ N_{R}+1/2}  \, ,
\end{equation}
from which we conclude, together with Eq.~(\ref{eq:accu}), that
\begin{eqnarray}\label{eq:mup_avg}
\langle \mu_P^{(\beta)} \rangle_{\rm RMT} & = & 
\langle \mu_{\rm tot}^{(\beta)} \rangle_{\rm RMT} = 0 \, .
\end{eqnarray}
RMT thus predicts the vanishing of the spin polarization on 
average. The latter is understood in the standard mesoscopic 
sense, as an average over changes in the dot shape, the overall
chemical potential or the position of the leads. 

Chemical potentials nevertheless 
exhibit spatial fluctuations in a single sample,
whose magnitude can
be estimated by calculating ${\rm var} \, \mu_P^{(\beta)}$.
Of further interest are also
mesoscopic fluctuations of the sample-integrated chemical potentials
whose magnitude we analyze with 
${\rm var} \, \mu_{\rm tot}^{(\beta)}$. 
To calculate the variance of $\mu_P^{(\beta)}$ in Eq.~(\ref{eq:accu}), we 
take the variance of the numerator and divide it by the
squared average of the denominator, a procedure which neglects
subdominant corrections ${\cal O}(N_{L,R}^{-1})$.
The variance of the spin-dependent transmission coefficients have been
calculated in Ref.~\cite{Bar07}. For $i =L,R$, we have
\begin{subequations}
\begin{eqnarray}
{\rm var} \mathcal{T}^{(\beta 0)}_{Pi}= \frac{4 [ N_P N_i (N_{\rm T}-1) - N_P N_i^2]}{N_{\rm T} (2 N_{\rm T}-1) (2 N_{\rm T}-3)} \, ,\\ 
\label{eq:covar}
{\rm covar} (\mathcal{T}^{(\beta 0)}_{PL},
\mathcal{T}^{(\beta 0)}_{PR} )= \frac{-4 N_P N_L N_R}{N_{\rm T} (2 N_{\rm T}-1) (2 N_{\rm T}-3)} \, ,
\end{eqnarray}
\end{subequations}
with $N_{\rm T}=N_L+N_R+N_P$. For our case with $N_P=1$, we obtain
\begin{eqnarray}\label{eq:var_mup}
{\rm var} \, \mu_P^{(\beta)} & = &16 \, \left(\frac{e V}{4} \right)^2 
\frac{N_L N_R}{N_{\rm T}
(2 N_{\rm T}-1) (2 N_{\rm T}-3)} \, .\qquad
\end{eqnarray}
For $N_L=N_R\equiv N \gg 1$, Eq.~(\ref{eq:var_mup}) goes into 
Eq.(\ref{eq:varmup_N}).
 
With the further assumptions, discussed
in the previous section, that (i) spin relaxation processes are frozen in 
the probe and (ii)
the correlator $
\langle \mathcal{T}^{(\alpha 
\beta)}_{PL} \mathcal{T}^{(\alpha \beta)}_{P'L} \rangle
- \langle  \mathcal{T}^{(\alpha 
\beta)}_{PL} \rangle \langle\mathcal{T}^{(\alpha \beta)}_{P'L} \rangle $
sharply decays when the distance between the probe positions
$P$ and $P'$ reaches the Fermi wavelength, we get the typical total
spin-dependent chemical potentials
\begin{equation}\label{eq:var_tot}
{\rm var} \, \mu_{\rm tot}^{(\beta)} = 
\frac{k_{\rm F}^2 A}{4 \pi} {\rm var} \, \mu_P^{(\beta)}  \, .
\end{equation}
As justified above, we identify these potentials with those on the dot.
Assuming that they are generated by a spin polarization, we
convert them into the variance of the spin polarization by dividing them with
$2 \delta$. For 
$N_L=N_R\equiv N$, we obtain the variance of the total number of 
polarized spins on the dot as
\begin{eqnarray}\label{eq:varS_tot}
{\rm var} \, S_{\rm dot}^{(\beta)} = \frac{k_{\rm F}^2 A}{4 \pi}
\left(\frac{e V}{2 \delta} \right)^2 
\frac{N^2}{(2N+1)
(4 N+1) (4N-1)} 
\end{eqnarray}
which, for $N \rightarrow \infty$, gives Eq.~(\ref{eq:typaccu}).
The validity of Eqs.~(\ref{eq:var_mup}--\ref{eq:var_tot}) 
is numerically confirmed below in Fig.~\ref{fig:accu_tot}.

\begin{figure*}[t]
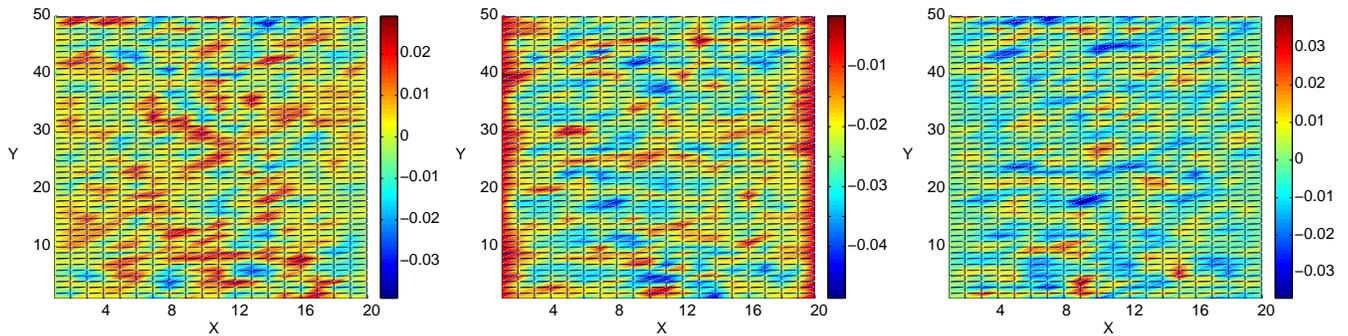

\includegraphics[width=0.32\textwidth]{fig2a.epsi}\hspace{1mm}
\includegraphics[angle=0,width=0.32\textwidth]{fig2b.epsi}\hspace{1mm}
\includegraphics[angle=0,width=0.32\textwidth]{fig2c.epsi}
\caption{\label{fig:local_accumul} Numerical proof-of-principle 
for the local measurement of spin accumulations 
on a single, diffusive, Rashba-coupled
lattice of size 20 $\times$ 50 traversed by an electric current in 
the positive $x$-direction (from left to right).
The on-site disorder and Rashba spin-orbit 
interaction correspond to a mean free path
$\ell=6a$ and a spin-orbit length $l_{\rm so}=9a$. 
The system has hard-wall boundary conditions
at its top and bottom boundaries. 
From left to right, spin accumulations 
$\mu^{(\beta)}$ with $\beta=x$, $y$ and $z$. 
Only $\mu^{(y)}$ displays a finite
average magnetoelectric effect (central panel), 
as predicted in Ref.~\cite{Ede90}.
Fluctuations exist for all spin orientations. 
The color scale is in units of the voltage bias $eV$.
Note the different color scales in different panels.}
\end{figure*} 

The power of our scattering theory is perhaps even more evident
when one realizes that it allows to calculate correlation
functions between local spin polarizations inside the sample
and spin currents in the leads. The spin current in lead $L$
is given by
\begin{eqnarray}
I_L^{(\alpha)} = \frac{e^2 V}{h} [\mathcal{T}_{LR}^{(\alpha 0)}-
\mathcal{T}_{LL}^{(\alpha 0)}]-\frac{e}{h} \sum_\beta 
\mathcal{T}_{LP}^{(\alpha \beta)} \mu_P^{(\beta)} \, ,
\end{eqnarray}
and each of these three terms gives a separate
contribution to 
the correlator $\langle I_L^{(\alpha)} \mu_P^{(\beta)} 
\rangle-\langle I_L^{(\alpha)}\rangle \langle \mu_P^{(\beta)} 
\rangle$. For $\alpha=x,y,z$,
the contribution of the third term vanishes because 
it contains the average of three spin-dependent transmission coefficients.
Furthermore, RMT gives 
${\rm covar} (\mathcal{T}^{(\alpha 0)}_{Pi}, \mathcal{T}^{(\beta 0)}_{jk}) 
\propto \delta_{ik} \delta_{\alpha\beta} \delta_{\alpha0}$ if $j \ne P$.
Therefore, for $\alpha=x,y,z$, 
the first two terms do not contribute either, unless
the probe is placed in the immediate vicinity, i.e. within a single
Fermi wavelength of the contact to the $L$ lead. In this case, 
nonvanishing covariances appear in 
$\langle \langle I_L^{(\alpha)} \mu_P^{(\beta)} \rangle \rangle$, 
see Eq.~(\ref{eq:covar}), and one obtains
\begin{equation}
\langle \langle I_L^{(\alpha)} \mu_P^{(\beta)} \rangle \rangle 
= 
\frac{-8 \, \delta_{\alpha \beta} \delta_{P x_L} 
\, N_L N_R}{N_T (2N_T-1) (2N_T-3)} \frac{e^3 V^2}{h} \, ,
\end{equation}
where $\delta_{P x_L}$ requires that the probe be placed within 
$k_{\rm F}^{-1}$ of the boundary between the SOI-coupled dot and the
lead. Summing over all possible probe positions, one obtains a
universal correlator $\langle \langle I_L^{(\alpha)} \mu_{\rm tot} ^{(\beta)} \rangle \rangle = {\cal O}(N^0)$ for $N_{L,R} \sim N \gg 1$, which
is consistent with the universality of spin currents 
fluctuations~\cite{Bar07,Naz07,Kri08,Ada09}.
RMT thus predicts that coherent polarizations
correlate with the spin currents only if they are right at the sample exit.

\section{Numerics}\label{section:numerics}

We numerically illustrate the validity of our scattering theory
of spin polarizations and confirm our main prediction, Eqs.~(\ref{eq:var_mup}-\ref{eq:var_tot}).
We use tight-binding models defined by the Hamiltonian
\begin{eqnarray}
{\cal H} = \sum_{i,j} \sum_{\alpha,s,s'} h^{(\alpha)}_{ij} 
\sigma^{\alpha}_{s,s'} c^\dagger_{is} c_{js'} \, ,
\end{eqnarray}
where the operators $c_{is}^\dagger$ ($c_{is}$) create (destroy) 
a spin-$1/2$ fermion with spin $s=\uparrow,\downarrow$
and $\sigma^{\alpha}_{s,s'}$ indicate elements of Pauli matrices.
To give a proof-of-principle for our approach, we consider a diffusive
model, while quantitative confirmations of our theory are 
solely based on the RMT model. 

For the diffusive model, 
$i=(i_x,i_y)$ labels a site on a two-dimensional
lattice of size $L_x \times L_y = (N_x a) \times (N_y a)$
with spacing $a$, giving the size $M = N_x \times N_y$
of the Hamiltonian matrix.
We further take
$h^{(\alpha)}_{ij} = i t_{\rm so}
\delta_{\alpha,x} \delta_{i_x,j_x} (\delta_{i_y+a,j_y}
+\delta_{i_y-a,j_y}) -i  t_{\rm so} \delta_{\alpha,y} \delta_{i_y,j_y}
(\delta_{i_x+a,j_x}
+\delta_{i_x-a,j_x}) - t \delta_{\alpha,0}[\delta_{i_x,j_x}
(\delta_{i_y+a,j_y}
+\delta_{i_y-a,j_y}) + \delta_{i_y,j_y}(\delta_{i_x+a,j_x}
+\delta_{i_x-a,j_x})] + w_i \delta_{\alpha,0} \delta_{i,j}$, corresponding to 
an on-site disorder potential with randomly distributed 
$w_i \in [-w/2,w/2]$, a nearest-neighbor
hopping $t$ and a Rashba spin-orbit interaction $t_{\rm so}$. These parameters
give a spin-orbit length
$\ell_{\rm so} \approx \pi a t/t_{\rm so}$ and an elastic mean
free path $\ell \approx 48 a t^2/w^2$ close to half filling. 

For the RMT
model, we take $\sum_\alpha h^{(\alpha)}_{ij} \sigma^{\alpha}$ to be randomly 
distributed quaternions~\cite{Bro96,mehta}, 
with time-reversal symmetry requiring
$h^{(0)}_{ij}=h^{(0)}_{ji}$, and
$h^{(\alpha)}_{ij}=-h^{(\alpha)}_{ji}$ for $\alpha =x,y,z$. 
Generally, the indices $i,j$ have no particular physical meaning in RMT
and the size $M$ of the Hamiltonian matrix 
is a free parameter. Here we attach the system to external leads,
and also consider $i$ and $j$ to label sites on a two-dimensional
lattice whose Hamiltonian has an infinite-range random hopping. 
Once the sharp decay of spin-spin correlations is confirmed, $M= k_{\rm F}^2 A/4 
\pi$ gives the total number
of uncorrelated probe positions. 
For both models, all our simulations fix the Fermi energy 
slightly below half-filling.

To construct the scattering matrix, we follow 
the method of Ref.~\cite{Heidelberg}. 
We introduce a $M \times N_{\rm T}$ projection matrix $W$ which connects
the system to external leads. This matrix has only one nonzero element per
row, at a spot determined by the coupling between the system and
the leads.
For an ideal contact, all nonzero matrix elements of $W$ are unity.
The $N_{\rm T} \times N_{\rm T}$ scattering 
matrix
${\cal S}$ associated with ${\cal H}$ is given by
\begin{eqnarray}\label{eq:heidel}
{\cal S} = 1 - 2 \pi i W^\dagger (E_{\rm F} -{\cal H} + i \pi W W^\dagger)^{-1}
W \, ,
\end{eqnarray}
with the overall Fermi energy $E_{\rm F}$. 
The transmission coefficients of Eq.~(\ref{tr_coeff_spin}) 
are straightforwardly
obtained from ${\cal S}$.
For the diffusive model, this method is certainly not the most 
efficient numerically. We still rely on it, because for the RMT model,
the infinite-range hopping renders
the implementation of recursive Green's function algorithms 
problematic. Note that it is necessary to first generate
random Hamiltonians from which random scattering matrices
are constructed in order to consistently define probe
positions and investigate whether spin-spin
correlations decay sharply beyond a Fermi wavelength. To investigate the decay of these
correlations, we generate a random ${\cal H}$ and take 
\begin{eqnarray}\label{eq:W}
W & = & \left(
\begin{array}{ccc}
{\mathbb I}_{N_L} & 0 & 0\\
0 & \vec{v}_{M-N_L-N_R} & 0  \\
0 & 0 & {\mathbb I}_{N_R}
\end{array}
\right) \, ,
\end{eqnarray}
where for each choice of $W$, only a single component of the 
column vector $\vec{v}$ (whose index indicates its length)
is nonzero and equal to one. The position of that
nonzero component determines the position of the probe on the 
two-dimensional lattice. 
Our Hamiltonians are defined on rectangular two-dimensional lattices,
and we contact the transport modes at two of its edges, which is represented by
the two identity matrices in Eq.~(\ref{eq:W}).

We first show numerical proof-of-principle 
for the spin-probe mode. In Fig.~\ref{fig:local_accumul}, we 
present data for the local spin polarizations $\mu_P^{(\beta)}$, $\beta=x,y,z$.
The three panels clearly show that,
for transport in the $x$-direction and a Rashba SOI, 
there is a finite average magnetoelectrically induced polarization along the
$y$-axis only, as predicted in Ref.~\cite{Ede90}. Furthermore, 
there are local (as well as sample-to-sample)
fluctuations of the spin polarization in all directions, as 
predicted in Ref.~\cite{Duc08}. 
All accumulations vanish on average in the RMT model, moreover spatial
fluctuations occur on the smallest possible scale of a single site, as
expected, and we therefore do not show them here. 

\begin{figure}[t]
\includegraphics[angle=0,width=0.85\columnwidth]{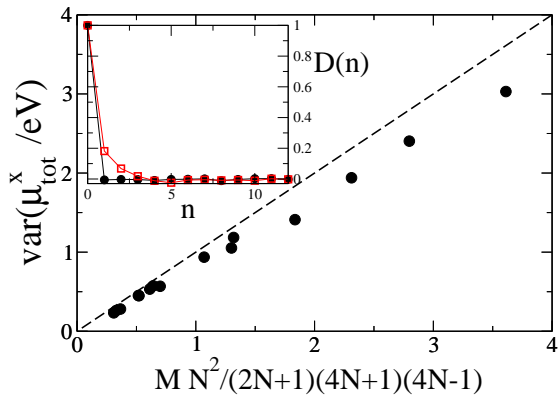}\\
\caption{\label{fig:accu_tot} Main panel: numerically obtained
variance of the sample-integrated spin-dependent potential $\mu_{\rm tot}^{(x)}$
in the RMT model (circles) 
vs. the theoretical prediction of Eq.~(\ref{eq:varS_tot}) (dashed line)
for $N_L=N_R=N \in [3,25]$
and $M \in [64,625]$. Deviations from the theory occur at
smaller number of channels, for which fluctuations in 
the denominator of Eq.~(\ref{eq:accu}) can no longer be neglected.
The behavior of $\mu_{\rm tot}^{(y)}$ and $\mu_{\rm tot}^{(z)}$ is
the same as the one of $\mu_{\rm tot}^{(x)}$. 
Inset: spin-spin correlation function 
$D(n)=\langle S^{(\beta)}(i) S^{(\beta)}(i+n  a) \rangle$ for 
the RMT  model (black circles) and the diffusive model (red squares) 
with $\ell=8 a$, $\ell_{so}=12 a$. In both cases, we find that the correlator
does not depend on the spin orientation $\beta$.
}
\end{figure}

We next confirm our conjecture of a sharp decay of correlations, and to that
end, we calculate the spin-spin correlation function
$D(n)=\langle S^{(\beta)}(i) S^{(\beta)}(i+n  a \vec{e}) \rangle$.
For both diffusive as well as RMT models we found no 
significant anisotropy, i.e. $D$ was the same for $\vec{e}$ in the $x$ or
$y$ direction. This is of course expected for the RMT model, but 
would deserve 
some more investigations in the diffusive case.
The Fermi wavelength being equal to the lattice spacing 
close to half filling, one expects
$D(n)=\delta_{n0}$ for the RMT model. This is 
confirmed in the inset of Fig.~\ref{fig:accu_tot}, where we additionally
see that $D(n)$ is also sharply peaked at $n=0$ in 
the diffusive model, with some correlation persisting over few, $\sim$2-3
sites. This sharp decay of $D(n)$ in the 
diffusive model suggests that our
conjecture of a fast decay of the correlator
$\langle \mathcal{T}^{(\alpha 
\beta)}_{PL} \mathcal{T}^{(\alpha \beta)}_{P'L} \rangle-  
\langle \mathcal{T}^{(\alpha \beta)}_{PL} \rangle \langle 
\mathcal{T}^{(\alpha \beta)}_{P'L} \rangle$ applies beyond RMT.
This also deserves further investigations.

Finally focusing on the RMT model,  Fig.~\ref{fig:accu_tot} confirms our main results, 
Eq.~(\ref{eq:var_mup}-\ref{eq:var_tot}), by showing the
scaling of the variance of the sample-integrated spin-dependent chemical potential
in symmetric samples $N_L=N_R=N$ as a function of $N$ and the size
$M$ of the random matrix Hamiltonian. Small deviations from our analytical
predictions occur at small $N$ for which our procedure of replacing
the denominator in Eq.~(\ref{eq:accu}) by its average squared when calculating
${\rm var} \, \mu_P^{(\beta)}$ is no longer justified. 
All these results 
numerically confirm our scattering theory of spin polarizations.

\section{Conclusions}

We have demonstrated how introducing an external probe allows to 
relate local current-induced 
spin polarizations in a mesoscopic system to spin-resolved
transmission coefficients. Focusing on ballistic systems, we used
random matrix theory to compute the statistics of these polarizations.
We predicted that a ballistic quantum dot with 
large contacts to external transport leads -- but still large dwell time -- 
typically
carries a spin polarization that exceeds those previously investigated
by a parametrically large factor [see Eqs.~(\ref{eq:typaccu}--\ref{eq:DL})].
Perhaps most importantly, we used this spin-probe model to calculate for
the first time quantum correlations between local spin polarizations
and spin currents, beyond the classical Fick law. 
We showed that the reported large spin polarizations correlate
only weakly with spin currents, thus they cannot be easily converted
into spin currents -- and vice-versa. As disappointing as this result may be,
it is consistent with the universality of spin current 
fluctuations~\cite{Bar07,Naz07,Kri08,Ada09}.

The reported accumulations are coherent in nature, and one might wonder
how they disappear as the system is coupled to external sources of
noise. In this respect it is interesting to note that for
a multichannel probe,
Eq.~(\ref{eq:varS_tot}) becomes, for
$N,N_P \gg 1$,
\begin{eqnarray}
{\rm var} \, S_{\rm tot}^{(\beta)} &\approx & \frac{k_{\rm F}^2 A}{144 \pi N}
\left(\frac{e V}{2 \delta} \right)^2 \frac{1}{(1+N_P/2N)^3}\, . 
\end{eqnarray}
Enlarging the spin-probe damps the variance of the polarizations
in a similar way as a dephasing probe damps conductance 
fluctuations~\cite{But}, with however a different parametric 
dependence on $N_P/N$. Note that the probe generating this damping is 
spin-conserving on time-average, as there is no spin current 
flowing through it. 

Beside decoherence effects, there are a number of interesting open problems, 
related for instance to spatial fluctuations of spin polarizations and 
our assumption of a sharp decay of correlations at $k_{\rm F}^{-1}$. Finding
out how the polarizations correlate to spin currents
in diffusive systems, as well as in true ballistic systems treated beyond 
RMT is also of interest. Ways to 
extract the large typical polarizations we reported in the nonlinear
transport regime or by optical means as in Ref.~\cite{Kat04}
should also be explored.

\section*{Acknowledgments}

This work has been supported by the NSF under grant DMR-0706319. 
We acknowledge the hospitality of 
the Basel Center for Quantum Computing and Quantum Coherence
at the early stage of this project. We thank 
M. Duckheim and D. Loss for an introduction to magnetoelectric effects
and discussions of their letter, Ref.~\cite{Duc08}, and M. B\"uttiker
for discussions on probe physics.

\section*{Appendix A}

We demonstrate that the potentials ensuring
zero current through a single-channel probe do not depend on the
transparency of the contact between the probe and the system.
According to Eq.~(\ref{eq:accu}),
the chemical potentials in an external probe connected to a voltage-biased 
two-terminal cavity are
given by the ratio of the difference and the sum of transmission
coefficients between the transport leads and the probe. We show that 
the $\Gamma_P$-dependence of the coefficients 
$\mathcal{T}_{PI}^{(\alpha \beta)}$ factorizes when $N_P=1$
so that $\Gamma_P$ does not affect the chemical potential.

Our starting point is Eq.~(\ref{eq:heidel}) for the scattering matrix,
which we rewrite here, 
\begin{eqnarray}
{\cal S} = 1 - 2 \pi i W^\dagger (E_{\rm F} -{\cal H} + i \pi W W^\dagger)^{-1}
W \, .
\end{eqnarray}
The matrix $W W^\dagger=V V^\dagger+P P^\dagger$ projects onto the 
system's states coupled to the
transport leads ($V V^\dagger$) and the probe ($P P^\dagger$).
We can write a perturbative series in $PP^\dagger$ for the 
elements of the scattering matrix connecting the leads to the probe 
($j_I$ is a channel index on the $I=L,R$ leads)
\begin{widetext}
\begin{eqnarray}\label{expansion}
{\cal S}_{P,j_I} & = & -2 \pi i \sum_{x,x'} P^*_{xP} V_{x' j_I}
\langle x | (E_{\rm F} -{\cal H} + i \pi W W^\dagger)^{-1} | x' \rangle \nonumber \\
&=& -2 \pi i \sum_{x,x'} P^*_{xP} V_{x' j_I} \left[
\langle x | (E_{\rm F} -{\cal H} + i \pi V V^\dagger)^{-1} | x' \rangle 
\right. \nonumber \\
&&\left. -i \pi \langle x | 
(E_{\rm F} -{\cal H} + i \pi V V^\dagger)^{-1} P P^\dagger 
(E_{\rm F} -{\cal H} + i \pi V V^\dagger)^{-1} | x' \rangle + \ldots \right] \, .
\end{eqnarray}
The transparency of the contact between probe and system is contained 
exclusively in the matrix $P$. We are free to choose any orthogonal 
basis $\{ | x\rangle \}$ for the system and pick one where the 
$M \times 1$ coupling 
matrix between probe and system is $P_{xP} = \omega_P \delta_{xP}$.
One then has for the $M \times M$ matrix $\langle x| PP^\dagger | x' \rangle  = 
\langle P| P P^\dagger | P \rangle \delta_{xP} \delta_{x'P} = 
|\omega_P|^2 \delta_{xP} \delta_{x'P}$, with $|\omega_P|^2 = (M \delta/\pi^2 \Gamma_P)
(2-\Gamma_P - 2 \sqrt{1-\Gamma_P})$~\cite{caveat3}.
One gets
\begin{eqnarray}
{\cal S}_{Pj_I} & = & -2 \pi i \omega_p^* \sum_{n=0}^\infty 
\left(-i \pi |\omega_P|^2 \, 
 \langle P|(E_{\rm F} -{\cal H} + i \pi V V^\dagger)^{-1} 
|P\rangle \right)^n \times 
\sum_{x'} V_{x' j_I} \langle P| 
(E_{\rm F} -{\cal H} + i \pi V V^\dagger)^{-1} |x'\rangle  \nonumber \\
&=& f(\Gamma_P) \times 
\sum_{x'} V_{x' j_I} \langle P| 
(E_{\rm F} -{\cal H} + i \pi V V^\dagger)^{-1} |x'\rangle \, .
\end{eqnarray}
\end{widetext}
One sees that, because the probe is single-channeled, the 
$\Gamma_P$-dependent 
part of ${\cal S}_{Pj_I}$ factorizes, 
and the same obviously holds for transmission
coefficients ${\mathcal T}_{PI}^{(00)} = \sum_{j_I} |{\cal S}_{Pj_I}|^2$. This is not the case for a multichannel probe,
in which case $| P \rangle$ is no longer unique.
Thus, for $N_P=1$, the electrochemical potential
in Eq.~(\ref{eq:accu_charge}) do not depend on the strength of the tunnel 
barrier. The argument is straightforwardly extended to the spin-dependent
chemical potentials in Eq.~(\ref{eq:accu}), 
because for a single-channel probe, the reflection
coefficient ${\mathcal T}_{PP}^{(\alpha \beta)}$ is diagonal in 
spin-indices.


\begin{thebibliography}{99}
\bibitem{cisp} F.T. Vasko and N.A. Prima, 
Sov. Phys. Solid State {\bf 21}, 994 (1979); L.S. Levitov, Y.V. 
Nazarov, and G.M. Eliashberg, 
Sov. Phys. JETP {\bf 61}, 133 (1985); A.G. Aronov and Yu.B. Lyanda-Geller, 
JETP Lett. {\bf 50}, 431 (1989).

\bibitem{Ede90} V.~M. Edelstein,
Solid State Comm. \textbf{73}, 233 (1990).

\bibitem{Dya71a} M.I. Dyakonov and V.I. Perel, 
Sov. Phys. JETP Lett. 
\textbf{13}, 467 (1971).

\bi{she} For a review on the spin Hall effect, see:
H.-A. Engel, E.I. Rashba, and B.I. Halperin, 
{\it Theory of Spin Hall Effects in Semiconductors}, 
Handbook of Magnetism and Advanced Magnetic Materials, 
Wiley and Sons (2007); arXiv:cond-mat/0603306.

\bibitem {Kat04}
Y.K. Kato, R.C. Myers, A. C. Gossard, and D. D. Awschalom,
Science \textbf{306}, 1910 (2004).

\bibitem{Wun05}
J. Wunderlich, B. K\"{a}stner, J. Sinova, and T. Jungwirth,
Phys. Rev. Lett. \textbf{94}, 047204 (2005).

\bibitem{Val06}
  S.O.~Valenzuela and M. Tinkham, 
Nature \textbf{442}, 176 (2006).

\bibitem {Sai06}
  E. Saitoh, M. Ueda, H. Miyajima, and G. Tatara,
Appl. Phys. Lett. \textbf{88}, 182509 (2006);
  T.~Kimura, Y. Otani, T. Sato, S. Takahashi, and S. Maekawa,
Phys. Rev. Lett. {\bf 98}, 156601 (2007); {\bf 98}, 249901(E) (2007).

\bibitem{Zha06} H. Zhao, E.J. Loren, H.M. van Driel, and 
A.L. Smirl, 
Phys. Rev. Lett. 96, 246601 (2006).

\bi{Mis04} E.G. Mishchenko, A.V. Shytov, and B.I. Halperin,
Phys. Rev. Lett. {\bf 93}, 226602 (2004).

\bi{Bur04} A.A. Burkov, A.S. N\'u$\tilde{\rm n}$ez, and 
A.H. MacDonald, Phys. Rev. B {\bf 70} 155308 (2004).

\bi{Mal05} A.G. Mal'shukov, L.Y. Wang, C.S. Chu, and K.A. Chao,
Phys. Rev. Lett. {\bf 95}, 146601 (2005).

\bibitem{Ada07} \.{I}. Adagideli, M. Scheid, M. Wimmer, G.E.W. Bauer, and K. 
Richter, 
New J. Phys. {\bf 9}, 382 (2007).

\bi{Tserk} Y. Tserkovnyak, B.I. Halperin, A.A. Kovalev, and A.
Brataas, 
Phys. Rev. B {\bf 76}, 085319 (2007).

\bi{Cha05} O. Chalaev and D. Loss, Phys. Rev. B {\bf 71}, 245318 (2005).

\bibitem{Bar07} J.~H. Bardarson, \.I.~Adagideli, and Ph. Jacquod,
Phys. Rev. Lett. {\bf 98}, 196601 (2007). 

\bibitem{Naz07} Y.~V. Nazarov, 
New J. Phys. {\bf 9}, 352 (2007).

\bibitem{Kri08}
J.~J. Krich and B.~I. Halperin, Phys. Rev. B {\bf 78}, 035338 (2008).

\bi{Ada09} \.I. Adagideli, J.H. Bardarson, and Ph. Jacquod, 
J. Phys.: Condens. Matter {\bf 21}, 155503 (2009).

\bi{Kri09} J.J. Krich, Phys. Rev. B {\bf 80}, 245313 (2009).


\bibitem{Nik05} B.K. Nikoli\'c, L.P. Z\^arbo, and S. Souma, 
Phys. Rev. B {\bf 72}, 075361 (2005).

\bibitem{guo} W. Ren, Z. Qiao, J. Wang, Q. Sun, and H. Guo,
Phys. Rev. Lett. {\bf 97}, 066603 (2006).

\bibitem{Duc08} M. Duckheim and D. Loss, 
Phys. Rev. Lett. {\bf 101}, 226602 (2008).


\bi{Bro96} P.W. Brouwer and C.W.J. Beenakker, J. Math. Phys. {\bf 37}, 
4904 (1996).

\bi{mehta} M.L. Mehta, {\it Random Matrices}, Academic Press, London (1990).

\bi{caveat1} Because of Coulomb interactions, such a direct relation
does not necessarily exist for the electrochemical potential on the dot,
see e.g.: M. B\"uttiker, J. Phys.: Condens. Matter {\bf 5}, 9361 (1993). 

\bibitem{But86} M. B\"uttiker, Phys. Rev. Lett. {\bf 57}, 1761 (1986).

\bibitem{Imr86} Y. Imry in {\it Directions in Condensed Matter Physics}, 
G. Grinstein and G. Mazenko eds., World Scientific, Singapore (1986).

\bi{caveat2} For $\beta=x,y,z$, $\mu_i^{(\beta)}$
gives the chemical potential difference between the two possible 
spin orientations along the corresponding spin axis,
while $\mu_i^{(0)} = E_F+e V_i$ gives the usual electrochemical
potential.
We refer to electrochemical and spin-dependent
potentials collectively as ``chemical potentials''.
Note the plural. ``Chemical potential'' still retains its standard
meaning.

\bi{symmetry_gs0} F. Zhai and H.Q. Xu, 
Phys. Rev. Lett. {\bf 94}, 246601 (2005);
A.A. Kiselev and K.W. Kim, Phys. Rev. B {\bf 71}, 153315 (2005).

\bibitem{But89} M. B\"uttiker, Phys. Rev. B {\bf 40}, 3409 (1989).

\bi{Jac06} Correlated escape of pairs of classical trajectories and its
influence on semiclassical transport was first discussed in: 
Ph. Jacquod and R.S. Whitney, Phys. Rev. B.  {\bf 73} , 195115 (2006). The only
nonvanishing semiclassical contributions to $\langle \mathcal{T}_{ij}^{(\alpha 0)} 
\mathcal{T}_{kl}^{(\alpha 0)} \rangle- \langle \mathcal{T}_{ij}^{(\alpha 0)} \rangle
\langle \mathcal{T}_{kl}^{(\alpha 0)} \rangle$ for $\alpha \ne 0$ 
have trajectories with correlated escape; Ph. Jacquod, unpublished (2008).

\bi{Heidelberg} C. Mahaux and H.A. Weidenm\"uller, 
{\it Shell-Model Approach to Nuclear Reactions}, North-Holland,
Amsterdam (1969).

\bibitem{But} M. B\"uttiker, Phys. Rev. B {\bf 33}, 3020 (1986); H.U. Baranger
and P.A. Mello, Waves in Random Media {\bf 9}, 105 (1999).

\bibitem{caveat3} The eigenvalue of $PP^\dagger$ 
$|\omega_P|^2 = (M \delta/\pi^2 \Gamma_P)
(2-\Gamma_P \pm 2 \sqrt{1-\Gamma_P})$ is not
uniquely determined by $\Gamma_P$. Here we take the smallest of the two
solutions, for which the expansion of Eq.~(\ref{expansion}) is guaranteed
to converge.

\end{thebibliography}
\end{document}